\documentstyle[12pt]{article}
\begin{document}
\baselineskip=0.7cm
\newcommand{\EQ}{\begin{equation}}
\newcommand{\EN}{\end{equation}}
\newcommand{\EQA}{\begin{eqnarray}}
\newcommand{\EQN}{\end{eqnarray}}
\newcommand{\e}{{\rm e}}
\newcommand{\Sp}{{\rm Sp}}
\renewcommand{\theequation}{\arabic{section}.\arabic{equation}}
\newcommand{\Tr}{{\rm Tr}}
\renewcommand{\thesection}{\arabic{section}.}
\renewcommand{\thesubsection}{\arabic{section}.\arabic{subsection}}
\makeatletter
\def\section{\@startsection{section}{1}{\z@}{-3.5ex plus -1ex minus 
 -.2ex}{2.3ex plus .2ex}{\large}} 
\def\subsection{\@startsection{subsection}{2}{\z@}{-3.25ex plus -1ex minus 
 -.2ex}{1.5ex plus .2ex}{\normalsize\it}}
\def\appendix{
\par
\setcounter{section}{0}
\setcounter{subsection}{0}
\def\thesection{\Alph{section}}}
\makeatother
\def\thefootnote{\fnsymbol{footnote}}
\begin{flushright}
hep-th/9808188\\
UT-KOMABA/98-21\\
August 1998
\end{flushright}
\vspace{1cm}
\begin{center}
\Large
Equations of Motion and Galilei Invariance  \\
in D-Particle Dynamics

\vspace{1cm}
\normalsize
{\sc Yuji Okawa}
\footnote{
e-mail address:\ \ okawa@hep1.c.u-tokyo.ac.jp}
and 
{\sc Tamiaki Yoneya}
\footnote{
e-mail address:\ \ tam@hep1.c.u-tokyo.ac.jp}
\\
\vspace{0.3cm}
 {\it Institute of Physics\\
University of Tokyo, Komaba, Tokyo 153 }

\vspace{1.3cm}
Abstract\\

\end{center}
As a continuation of our previous work on 
the multi-body forces of D-particles in 
supergravity and Matrix theory, 
we investigate the problem of motion.  
We show that the scattering of D-particles 
including recoil derived in Matrix theory 
 is precisely reproduced by supergravity 
with the discrete light-cone prescription up to the 
second order in 11 dimensional Newton constant.  
An intimate connection of recoil and Galilei 
invariance in supergravity is pointed out and elucidated.  

\newpage
\section{Introduction}

Matrix theory \cite{bfss} as a candidate for the non-perturbative 
formulation of M-theory has passed 
many nontrivial checks. In particular, 
the correspondence with 11 dimensional supergravity 
may be regarded as  a cardinal test. 
If Matrix theory would be proven to be well defined 
non-perturbatively and to agree with  
supergravity in the low-energy limit, 
that would be convincing evidence for establishing  
Matrix-theory conjecture.   

In our previous work \cite{oy} which is 
hereafter referred to as I, we have presented a detailed 
comparison between Matrix theory and supergravity 
with respect to the multi-body interactions 
of D-particles. We have shown that the 
effective actions on both sides precisely agree 
up to 3-body interaction terms. 
The effective actions are determined under a particular 
 approximation which is common on both sides. 
Namely, we have completely neglected the 
recoil of D-particles and compared the actions 
with the criterion that they should give the 
same eikonal phase shift. 
On Matrix-theory side, this allows us to 
perform the simple loop calculation with 
a fixed background which corresponds to 
straight-line trajectories of D-particles. 
On supergravity side, on the other hand, 
the eikonal approximation amounts to the 
approximation that the energy-momentum 
tensor of D-particles is assumed to be that of 
free D-particles. In so doing, it was crucial that 
the recoil effect can be separated from the rest 
in the particular 
gauge we have adopted.  

However, it is important to remember that 
the eikonal approximation is not 
a systematic expansion scheme in any rigorous 
sense. For this reason, we must take into account 
the recoil effect to make the computations 
on both sides completely self-consistent and 
must check whether the agreement persists 
after that. The main purpose of the present note 
is  to present such a check as promised in I. 
It is also an 
important exercise to see explicitly how the 
general relativistic equation of motion, i.e. 
geodesic equation which 
is nothing but the integrability condition for the 
gravitational field equation, fits into the 
loop expansion of Matrix-theory calculations 
where we cannot see, at least directly, 
 the gravitational field degrees of 
freedom which emerge only as a loop effect. 
Furthermore, obtaining the effective action 
using the eikonal approximation implicitly assumes 
that there exists the effective action which is 
local with respect to time.  Therefore it is desirable 
to confirm directly that the recoil effect 
calculated from the equations of motion obtained by the 
variation of the effective action coincides with the 
contribution of the tadpole diagram of Matrix theory.  
These checks 
would provide firm evidence that Matrix theory 
indeed describes the dynamics of D-particles 
beyond their restricted properties associated with BPS 
or supersymmetric constraint. 

In section 2, we first explain the apparent violation 
of Galilei invariance in the eikonal approximation in 
supergravity in the simplest case of two clusters 
of coincident D-particles and its resolution 
after taking into account the recoil effect.  
We then extend the arguments to the multi D-particle system 
treated in I. 
The effective action 
including the 
recoil correction is formulated in section 3 and the precise 
agreement between supergravity and Matrix 
theory is demonstrated. Although the 
treatment of recoil in Matrix theory 
is more or less self-evident from the general 
formalism of effective actions, we feel that our 
discussion emphasizing its 
connection with general relativistic equations of motion 
 will be useful as a basis for future works toward  
further systematic comparisons of higher  order effects 
in supergravity and Matrix theory. 

\vspace{0.5cm}
\section{Galilei Invariance and Recoil  in Supergravity}
\subsection{System of two clusters of coincident D-particles}
Let us start from considering the simplest case of two 
clusters of coincident D-particles, 
assuming that one of them, source, 
with $p_-= N_1/R$  
 is much `heavier' ($N_1\gg N_2$) than the other, probe, with $p_-=N_2/R$. The source can then be treated to be at rest 
in the transverse space, and the effective action for the 
probe with (transverse) velocity $v$ in the lowest nontrivial approximation is given by
\footnote{
We use the same conventions as in I. }
\EQ
S=\int ds \Bigl[ 
{N_2\over 2R}v^2 + {15N_1N_2\over 16 M^9 R^3}
{v^4 \over r^7} 
\Bigr]  .
\label{effectiveaction1}
\EN   
Here the time $s$ is identified with the light-cone time in 
11 dimension as $s =x^+/2=(x^{11}+t)/2$. 
The second term is the gravitational energy 
of the probe corresponding to the 
one-graviton exchange contribution
\EQ
\int d^{11}x \,  {1\over 2}\zeta_{1\, \mu\nu}(x) \, 
\tau_2^{\mu\nu}(x) ,
\EN 
where $\tau_2^{\mu\nu}$ and $\zeta_{1\,  \mu\nu}$ 
are the lowest order energy-momentum tensor of the 
probe and the gravitational field produced by the 
fixed source, respectively, 
\EQ
\tau_2^{\mu\nu}(x)=
{N_2\over 2\pi R^2}\delta^9(x-x(s))s^{\mu}_2s_2^{\nu} ,
\label{energy-momentumlowest} 
\EN
\EQ
\zeta_{1\, \mu\nu}(x)=
{15\over (2\pi)^4}\kappa_{11}^2{N_1\over 2\pi R^2} 
{s_{1\, \mu}s_{1\, \nu} \over r^7} , 
\EN
with $r$ being the relative transverse distance between 
the source and probe. In our convention, the 11D Newton constant is 
$\kappa_{11}^2 = 16\pi^5 /M^9 \, \, 
(M^{-3}=g_s\ell_s^{3}) $ and the compactification 
radius along the $x^-= x^{11}-t$ direction is $R \,(=
g_s\ell_s)$. These expressions are 
obtained by averaging over the $x^-$ direction from 
$0$ to $2\pi R$. 
The velocity vectors $s^{\mu}
\equiv dx^{\mu}/ds $ are 
\EQ
(s_{1\, +}, s_{1\, -}, s_{1\, i}) = (0, 1, 0) , \quad 
(s_2^+, s_2^-, s_2^i) = (2, -{1\over 2} v^2, v^i) ,
\label{velocity1}
\EN 
for the source and probe, respectively. 

Now 
the variation of the above effective action gives 
the equation of motion for the 
probe
\EQ
{N_2\over R}{d v^i\over d s} +
{15N_1N_2\over 4M^9R^3}{d\over ds}
\Bigl({v^iv^2\over r^7}\Bigr)+
{105N_1N_2\over 16M^9R^3}{v^4r^i\over r^9}=0 .
\label{variationaleq1}
\EN
On the other hand, the equation of motion in General 
Relativity is the geodesic equation, 
which should be regarded as the integrability 
condition for the field equation. 
It is not entirely obvious whether the latter 
coincides with the former variational equation 
of motion, because we have derived the effective 
action in supergravity using the eikonal approximation. 

Since the above 
effective action or energy-momentum 
tensor corresponds to the following probe action 
\EQ
S_2 ={N_2\over 2R}\int \, ds \,   
g_{\mu\nu}(x(s) )\, 
{dx^{\mu}(s)\over ds}{dx^{\nu}(s) \over ds} ,
\label{probeaction}
\EN
with $g_{\mu\nu} =\eta_{\mu\nu}+ \zeta_{1\, \mu\nu}$, 
we should adopt the variational equation 
of this action as the geodesic equation. 
However, we find that the spatial component of the 
geodesic equation is then 
\EQ
{N_2\over R}
 {d^2x^i(s)\over ds^2}
-{N_2\over 2R}\partial_i \zeta_{1\, --} (x(s)) (s_2^-)^2
=0 ,
\label{geodesiceq1}
\EN
where use is made of the fact that only 
nonzero component of the gravitational 
field of the source D-particle system is $\zeta_{1\, --}$. 
Comparing with the variational equation of motion 
(\ref{variationaleq1}) derived from the effective action,  
the geodesic equation of motion contains only the 
first and the last terms. The second term of 
(\ref{variationaleq1}) 
which comes from the velocity dependence 
of the interaction is missing in (\ref{geodesiceq1}). 

Let us next consider the equation of motion in the 
different Galilean frame in which the probe is 
at rest in transverse space. Then the velocity vectors must 
be replaced by 
\EQ
(s_{1\, +}, s_{1\, -}, s_{1\, i}) =(-{1\over 4} v^2, 1, -v^i)  , \quad 
(s_2^+, s_2^-, s_2^i) = (2, 0, 0) .
\label{velocity2}
\EN 
The geodesic equation is now
\EQ
{N_2\over R}\Bigl(
 {d^2x^i(s)\over ds^2}
+2{d\zeta_{1\, i+}(x(s))\over ds}
\Bigr)
-{2N_2\over R}\partial_i \zeta_{1\, ++} (x(s)) 
 =0 ,
\EN
which coincides with (\ref{variationaleq1}) after substituting 
the velocity vectors (\ref{velocity2}). Note 
that in this case we have used the fact that only 
nonzero component of 
the velocity vector of the probe is $s_2^+=2$. 

Thus we found that the geodesic equation is 
not Galilei invariant in the above naive 
argument based on the eikonal approximation. 
The resolution of this apparent contradiction is 
as follows. Since the D-particle is assumed to be a Kaluza-Klein mode of the graviton, 
its trajectory must satisfy the massless condition, 
\EQ
g_{\mu\nu}(x(s) )\, 
{dx^{\mu}(s)\over ds}{dx^{\nu}(s) \over ds} =0 ,
\label{masslesscondition}
\EN
and the momentum constraint, 
\EQ
p_- ={N\over R} .
\EN
In reference \cite{bbpt}, the massless condition 
is imposed by taking the zero mass limit $m\rightarrow 0$ 
starting from the usual action 
$$S_{bbpt}=-m \int ds \sqrt{-g_{\mu\nu}(x(s) )\, 
{dx^{\mu}(s)\over ds}{dx^{\nu}(s) \over ds} 
} \, ,$$ and the momentum  constraint is taken into 
account by moving to the Routhian 
$$S_{bbpt}\rightarrow 
S_{bbpt}-\int ds \,  p_-{dx^-\over ds} . $$ 
In quantum theory, the transition to the Routhian is 
explained as the Fourier 
wave function corresponding to the change from 
  the coordinate to momentum representation.  

To deal with the above difficulty, however, 
it is more appropriate to 
start from the action with a Lagrange multiplier 
corresponding to the massless condition (\ref{masslesscondition}) 
as adopted in I, 
\EQ
S_D = {N\over 2R}\int \, ds \, \lambda(s)\,  
g_{\mu\nu}(x(s) )\, 
{dx^{\mu}(s)\over ds}{dx^{\nu}(s) \over ds}
-\int ds \, p_-{dx^-\over ds} ,
\label{routhian}
\EN
where $s$ is an arbitrary parametrization along the 
D-particle trajectory. 
The reparametrization invariance allows us to 
adopt the gauge $s = x^+/2$. 
Note that, by defining the Routhian, the independent 
variables in the longitudinal direction are 
$x^-$ and $p_-$, and the equation of motion for $x^-$ 
must be reinterpreted as the momentum constraint
\EQ
p_-={N\over R}\lambda(s) g_{\mu -}{d x^{\mu} \over ds}
={N\over R}g_{\mu -}{d x^{\mu} \over d\tau} ={N\over R}=constant ,
\label{momentumconstraint}
\EN
where we have set 
\EQ
d\tau = {ds \over \lambda(s)} .
\EN
The parameter $\tau$ is invariant under 
reparametrization of the D-particle trajectory. 
If the action and the energy-momentum tensor of 
D-particles 
are expressed in terms of $\tau$, we can eliminate the 
Lagrange multiplier $\lambda(s)$.   

Let us confirm the consistency of the above momentum 
constraint with the equations of motion and the 
field equation.  
In the classical approximation, we have to 
require that the variation of the Routhian 
with respect to $x^-$ keeping $p_-$ fixed must vanish 
without surface term.   
We then have, denoting the first term (integrand) 
of (\ref{routhian}) 
by $L$, 
\EQ
{\partial L\over \partial (dx^-(s)/ds)}\Big|_{x^-}
{\partial (dx^-(s)/ds)\over \partial x^-(s)}\Big|_{p_-}
+{\partial L\over \partial  x^-(s)}\Big|_{\dot{x}^-} 
-
p_-{\partial (dx^-(s)/ds)\over \partial x^-(s)}\Big|_{p_-}=0 ,
\EN
which, combined with (\ref{momentumconstraint}), 
leads to 
\EQ
{\partial L\over \partial  x^-(s)}\Big|_{\dot{x}^-}=0 . 
\label{pconstant}
\EN
On the other hand, the energy-momentum  tensor 
derived from the above action is 
\EQ
T^{\mu\lambda}(x) =
{N\over R} \int ds \, \lambda(s)
{dx^{\mu}(s)\over ds}{dx^{\lambda}(s) \over ds}
{1\over \sqrt{-g(x(s))}}\delta^{11} (x-x(s)) . 
\footnote{We note that, to the order we are interested in, 
the determinant of the space-time 
metric can always be treated as $-1$ because of the massless condition $s^{\mu}s_{\mu}=0$. }
\label{energymomentumtensor1}
\EN
The conservation, $D_{\mu}T^{\mu\nu}=0, $ of the energy-momentum tensor 
which is nothing but the integrability condition 
of the field equation requires that the 
trajectories of D-particles must satisfy the 
geodesic equation 
\EQ
{d^2 x^{\mu}\over d\tau^2} + \Gamma_{\alpha \beta}^{\mu}
{d x^{\alpha}\over d\tau}{d x^{\beta} \over d\tau} = 0 .
\label{geodesiceq}
\EN
The geodesic equation is equal to the variational 
equation derived by assuming the action consisting of 
only the first term in (\ref{routhian}). The 
equation (\ref{pconstant}) is then equivalent to 
$dp_-/ds=0$, as it should be to be compatible 
with the momentum constraint. 
By examining the compatibility of this 
condition and the geodesic equation (\ref{geodesiceq}), 
we find that the constancy of  $p_-$ is 
satisfied for arbitrary velocities of D-particle 
when 
\EQ
\partial_- g_{\mu\nu}=0  ,
\EN
which is again as it should be, since 
there is no exchange of $p_-$ momentum, and 
is consistent with the averaging prescription 
along the compactification circle 
in defining the energy-momentum tensor 
as in (\ref{energy-momentumlowest}).  Physically, this prescription 
is justified since $R$ is small for small 
$g_s$.  Note that the 
actual expansion parameter in our 
calculation is 
$\kappa_{11}^2 /R^2 \propto g_s^3/g_s^2=g_s$. 

Let us now consider the momentum constraint 
(\ref{momentumconstraint}).  
In our gauge $s=x^+/2$, it reduces to 
\EQ
1=\lambda(s)(1 + \zeta_{1\, \mu -}{dx^{\mu}\over ds})  . 
\label{pminusrelation}
\EN
This indicates that, once the geodesic 
equation of motion is taken into account,  
the choice $\lambda(s)=1$ 
  is allowed only when the relative velocity of D-particles are zero.  
In particular, for the probe system in the above 
discussion, the relation (\ref{pminusrelation}) 
becomes, to the first nontrivial order, 
\EQ
\lambda(s) = 1-\zeta_{1 \, \mu -}{dx^{\mu}\over ds}=
1 + {15N_1\over 4M^9R^2}{v^2 \over r^7} .
\label{lamdacorrection1}
\EN
It is easy to check that this also ensures the 
consistency of the 
 gauge condition $x^+=2s$ with the $+$ component 
of the geodesic equation (\ref{geodesiceq}).   
We also note that the geodesic equation 
is compatible with the massless condition 
(\ref{masslesscondition}). This ensures that the 
$x^-$ expressed in terms of the transverse components 
solving the massless constraint 
satisfies the geodesic equation automatically. 

The spatial component of the 
geodesic equation of the probe with the light-cone 
time $s=x^+/2$, is, in the present 
approximation, 
\EQ
{N_2\over R}
 {d\over ds}\Bigl((1+ {15N_1\over 4M^9R^2}{v^2 \over r^7}){dx^i(s)\over ds}\Bigr)
-{N_2\over 2R}\partial_i \zeta_{1\, --} (x(s)) (s_2^-)^2
=0 , 
\label{geodesiceq3}
\EN
coinciding with the variational equation of motion 
corresponding to the effective action (\ref{effectiveaction1}). 
Note that the shift of $\lambda(s)$ for the probe 
does not affect its Routhian, since $dx^-/ds$ is 
obtained by solving the massless constraint 
which does not contain the Lagrange multiplier 
directly. 
For the heavy source, the  correction to 
$\lambda(s)=1$ is proportional to $N_2 (\ll N_1$) 
and can be neglected compared with (\ref{lamdacorrection1}). 
Thus the shift of $\lambda(s)$ does not change the 
form of the effective action. However, 
Galilei invariance of the geodesic 
equation of motion on the side of 
supergravity and the agreement 
with the equation of motion 
derived from the effective action 
are recovered only by properly 
taking into account the 
Lagrange multiplier corresponding to the 
massless condition.  Galilei invariance is of course 
expected in the light-front formulation of 
supergravity adopting $x^+$ as the time. 

This is a part of the recoil effect which has 
been neglected in previous calculations \cite{bbpt}\cite{dr}
\cite{italy}\cite{taylor}\cite{suss}  
of D-particle interactions, including our work I. 
The eikonal approximation amounts to 
treating the D-particle trajectories 
as if they were completely independent 
of each other. Then the only 
natural choice of the energy-momentum tensor 
would be that of the free 
D-particle corresponding to the assumption 
$\lambda(s)=1 \, \, (\tau=s=x^+/2 )$ as has been 
adopted in I following \cite{bbpt}.  
Since the Lagrange multiplier $\lambda(s)$ essentially 
represents the arbitrariness of 
parametrization for the trajectories of D-particles, 
its shift means that the invariant parameter $\tau$ of each 
D-particle must be deviated from the external time 
parameter $x^+\, (=2s)$ depending on the relative 
motions of D-particles.  
The total recoil effect $\delta_{rc}x^i(s)$ 
obtained from the equation of motion 
(\ref{variationaleq1}) 
for the probe system 
can be expressed as the 
correction for  the velocity vector 
\EQ
{d \delta_{rc} x^i\over ds} = -
{15N_1\over 4M^9R^2}{v^2 \over r^7}v^i
-\int_{-\infty}^s ds'\, 
{105N_1\over 16M^9R^2}{v^4r^i\over r^9} ,
\EN
to the first order in $\kappa_{11}^2$.  
The first term in the right hand side can be 
interpreted as due to the change of parametrization 
\[
ds \rightarrow 
(1 + {15N_1\over 4M^9R^2}{v^2 \over r^7})ds ,
\]
which is required to keep the  gauge 
$x^+=2s$ after the recoil effect is taken into account.  
It is quite remarkable that the simple-looking 
effective action  and the associated 
equation of motion for D-particles automatically 
account for these subtle properties of 
D-particle motions in supergravity, in spite 
of the fact that the effective action itself 
was derived using the eikonal approximation. 

\subsection{Multi-cluster system} 
It is straightforward to extend our analysis 
to multi-cluster systems. The effective action
\footnote{
Here and in what follows we assume the convention that 
when the sums over particles hit the case $a=b$ leading 
vanishing relative distance $r_{ab}=0$, the 
contribution identically 
vanishes because of the vanishing relative 
velocity $v_{ab}\equiv 0$.  
We note that, in Matrix-theory calculations, 
there does not occur 
any infrared singularity associated 
with the massless (diagonal)  modes with this prescription. }  is 
\EQ
S_{eff}= \int ds\, \Bigl[
\sum_a{N_a \over R}
{1\over 2}\Bigl({dx_a^i \over ds}\Bigr)^2 
+ {1\over 2}\sum_{a, b} {15N_aN_b \over 16R^3 M^9}{v_{ab}^4 \over r_{ab}^7}  
+L_3 + \cdots  \, 
\Bigr] ,
\label{effectiveactiongeneral}
\EN
where $L_3$ is the second order term as derived in \cite{oy}
\EQ
L_3 = L_V +L_Y ,
\EN
\EQ
L_V=  -\sum_{a,b,c}{(15)^2 N_aN_bN_c\over 64 R^5M^{18}}
v_{ab}^2 v_{ca}^2 (v_{ca}\cdot v_{ab})
{1\over r_{ab}^7}{1\over  r_{ca}^7} ,
\label{lv}
\EN
\EQA
L_Y  &=&-\sum_{a, b, c} 
{(15)^3 N_aN_bN_c \over 24(2\pi)^4R^5M^{18}}
\Bigl[
-(s_b\cdot s_c) 
(s_{c}\cdot s_a)(s_b\cdot \tilde{\partial}_c)
(s_a\cdot \tilde{\partial}_{c})\nonumber \\
&&+{1\over 2}( s_{c}\cdot s_a)^2(s_b\cdot \tilde{\partial}_c)^2  
+{1\over 2}(s_b \cdot s_c)^2 (s_a\cdot \tilde{\partial}_{c})^2
\nonumber \\
&&-{1\over 2}(s_{b}\cdot s_a)
(s_a\cdot s_{c})(s_b \cdot \tilde{\partial}_c)(s_c \cdot\tilde{\partial}_b)
 \nonumber\\
&&+{1\over 4}(s_b\cdot s_c)^2 
(s_a\cdot \tilde{\partial}_{b})(s_a\cdot \tilde{\partial}_{c})
\Bigr]\Delta(a,b,c) , 
\EQN
where
\[
\Delta(a,b,c) \equiv
\int d^9 y {1\over |x_a-y|^7 |x_b -y|^7|x_c-y|^7} 
\]
\EQ
= {64(2\pi)^3 \over (15)^3}
\int_0^{\infty} d^3\sigma 
(\sigma_1\sigma_2 + \sigma_2\sigma_3 + \sigma_3\sigma_1)^{3/2}
 \exp \bigl( 
-\sigma_1|x_a-x_b|^2 -\sigma_2|x_b -x_c|^2 -\sigma_3|x_c-x_a|^2 \bigr)   ,
\label{delta} 
\EN
with $d^3 \sigma=d\sigma_1  d\sigma_2 d\sigma_3 $
and the notation $\tilde{\partial}$ is defined by
$\tilde{\partial}_{\mu}=(\partial_+, 0, -\partial_i)$.
As we will see below, to  compare with 
the available Matrix-theory calculations,  it is 
sufficient to take into account the recoil correction to 
the first nontrivial order, namely to the 
first order in 11D Newton constant $\kappa_{11}^2$. 

The D-particle part of the action is 
now 
\EQ
S_D = \sum_a{N_a\over 2R}\int \, ds \, \lambda_a(s)\,  
g_{\mu\nu}(x_a(s) )\, 
{dx_a^{\mu}(s)\over ds}{dx_a^{\nu}(s) \over ds}
-\sum_a\int ds p_{a\, -}{dx_a^-\over ds} .
\EN
 The eikonal approximation corresponds to 
the choice $\lambda_a(s)=1$ with straight-line 
trajectories $dx^i_a/ds =v^i_a$=constant in the 
gauge $s=\tau=x^+/2$.  
The momentum constraint is now 
\EQ
1=\lambda_a(s) g_{\mu -}(x_a){d x_a^{\mu} \over ds} .
 \EN
In the present approximation, the solution to this 
equation is given by
\EQ
\lambda_a(s) = 1 + \sum_b {15N_b \over 4M^9R^2} 
{v_{ab}^2 \over r_{ab}^7} ,
\label{lambda2}
\EN
where $v_{ab}$ and $r_{ab}$ are relative 
transverse velocity and distance, respectively. 
As in the previous case of two clusters, 
(\ref{lambda2}) ensures the consistency 
of the gauge condition with the geodesic 
equation of motion. 
The spatial component of the 
geodesic equation of motion to the 
first order in $\kappa_{11}^2$ is then 
\EQ
{N_a\over R}{d\over ds}\Bigl(v_a^i+\sum_b
 {15N_b\over 4M^9R^2}{v_{ab}^2 \over r_{ab}^7}v_{ab}^i\Bigr)
-{N_a\over 2R}\partial_i \zeta_{\mu\nu} (x_a(s)) s_a^{\mu}s_a^{\nu}
=0 , 
\label{geodesiceqgeneral}
\EN
where the gravitational field is 
\EQ
\zeta_{\mu\nu}(x_a)=\sum_b 
 {15N_b\over 2M^9 R^2} 
{s_{b\, \mu}s_{b\, \nu} \over r_{ab}^7} . 
\EN
It is easy to check that the geodesic equation 
(\ref{geodesiceqgeneral}) coincides with the 
variational equation derived from the effective 
action (\ref{effectiveactiongeneral}) in the 
present approximation.  We note that,  
if the deviation of the Lagrange multiplier $\lambda_a(x)$ 
from the eikonal approximation $\lambda_a(s)=1$ 
were neglected in supergravity, the second term in the parenthesis would  
have been 
$$-\sum_b{15N_b\over 4M^9R^2}{v_{ab}^2 \over r_{ab}^7}v_b^i ,$$ 
({\it i.e.} $v^i_{ab} \rightarrow -v^i_b$) which violates Galilei invariance. 

Furthermore, the shift of $\lambda_a(s)$ does not 
affect the form of the effective action itself (\ref{effectiveactiongeneral}),  
since the change of the two-body term induced by the 
shift of the Lagrange multiplier is canceled
\footnote{
This is easily seen by examining the 
equations (2.47) and (2.48) of I. The coefficients 
$-1/4$ and $1/2$, respectively, 
 in these expressions are responsible for the 
cancellation. Note that as already 
mentioned in the previous subsection, 
the form of the Routhian does not directly receive correction 
from $\lambda$.  
} between the two 
contributions coming from the pure gravity part and 
the D-particle part $\int ds \, p_-dx^-/ds$ 
in the present approximation.  This cancellation 
explains why the equation of motion obtained 
by the variation of the effective action in the eikonal approximation 
agrees with the geodesic equation. 
The total shift induced by recoil is thus determined 
only by the shift $\delta_{rc} x_a^i$ of the trajectories of D-particles 
as 
\EQ
{d \delta_{rc} x_a^i\over ds} = -
\sum_b
\Bigl( {15N_b\over 4M^9R^2}{v_{ab}^2 \over r_{ab}^7}
\, v_{ab}^i
+
\int_{-\infty}^s ds'{105N_b\over 16M^9R^2}{v_{ab}^4\over r_{ab}^9}
\, r_{ab}^i\Bigr) .
\label{shiftgeneral}
\EN

\vspace{0.5cm}
\section{Effective Action Including Recoil Corrections}
Our next task is to study the change of the value ({\it i.e.} 
scattering phase shift) of the effective action 
caused by recoil. 
Before going into the calculation, we have to 
pay attention to one subtlety associated with surface 
terms.  Namely, in computing the scattering phase shift, 
we have to take into account the wave function 
$\exp \Bigl(- i\sum_a p^i_a x^i_a\Big|^{t=\infty}_{t=-\infty}
\Bigl)$
of the initial and final states.  In classical approximation 
(or the lowest WKB approximation), the surface terms  
with $p_a^i =N_av_a^i/R$ are  
necessary in order to ensure that the first variation 
around the classical trajectories vanishes even at the 
asymptotic past and future. When we include the 
shift induced by recoil,  we have to assume that these 
surface terms are simultaneously shifted 
corresponding to the change of the initial or 
final states of D-particles.  Since the 
surface terms are arranged such that the 
surface contributions under the variation vanish, 
we can then safely neglect the 
surface contributions in evaluating the 
effect of recoil using partial integration. 
We note that this remark applies to both 
supergravity and Matrix theory calculations. 

\subsection{Recoil correction of the effective action}
Let us now proceed to calculate the   
effective action including recoil.  
We denote  the quantities before recoil correction by putting 
subscript $0$.  
Thus the trajectory is now $x^i_a(s) =
x_{a, 0}^i(s) + \delta x_a^i(s)$. 
We are interested in the effective action 
to the second order in Newton constant $\kappa_{11}^2$, 
or more appropriately, to the second order in $g_s$ 
comparing with the lowest order (which is 
of order $g_s^{-1}$) term, 
since the actual expansion parameter 
is $\kappa_{11}^2/R^2 \propto g_s$ because of the 
compactification along the $x^-$ direction. 
It is then sufficient to consider the equation of motion 
including only the first order interaction term.  
We denote the terms of the effective action 
by putting the superscript indicating the 
order with respect to the string coupling $g_s$. 
For notational simplicity, we set $\ell_s=1 
\, (M^{-3}=g_s, R=g_s) $. 
\EQ
S^{(-1)}= \int ds \sum_a{N_a\over 2g_s}v_a^2  ,
\EN 
\EQ
S^{(0)}={1\over 2}\int ds \sum_{a,b} {15N_aN_b \over 16}{v_{ab}^4 \over r_{ab}^7}. 
\EN
By substituting the trajectory with the recoil corrections, 
the effective action $S^{(-1)} + S^{(0)}$ becomes 
\[
S^{(-1)} + S^{(0)} 
\rightarrow 
\int ds \sum_a{N_a\over 2g_s}(v_{a,0} +\delta v_a)^2
+ {1\over 2}\int ds \sum_{a, b} {15N_aN_b \over 16}{(v_{ab,0}+\delta v_{ab})^4 \over (r_{ab,0}+\delta r_{ab})^7}
\]
\EQ
=S^{(-1)}_0 + S_0^{(0)} + S'^{(1)}
+ O(g_s^2) , 
\EN
where 
\EQ
S^{(-1)}_0=\int ds \sum_a{N_a\over 2g_s}v_{a,0}^2 ,
\EN
\EQ
S^{(0)}_0=
{1\over 2}\int ds \sum_{a,b} {15N_aN_b \over 16}{v_{ab,0}^4 \over r_{ab,0}^7} ,
\EN
\EQ
S'^{(1)}=-\int ds \sum_a{N_a\over 2g_s}(\delta v_{a})^2  .
\label{recoilcorrection}
\EN
In deriving this result, we have used the fact that 
$x^i_a(s) =
x_{a, 0}^i(s) + \delta x_a^i(s)$
is the solution of the variational equation of 
the effective action. The surface term 
in performing partial integration can be discarded as 
discussed above. 
The last term represents the recoil correction to 
the action and hence to the phase shift. 
If further partial integrations are allowed, 
we can formally rewrite this expression 
in terms of the recoil acceleration $\delta \alpha_a
=d\delta v_a/ds$ as 
\EQ
S'^{(1)}
={1\over 2}\sum_a N_a\int ds_1ds_2 \, \delta \alpha_a(s_1) 
D_x (s_1,s_2) \delta \alpha_a(s_2)  , 
 \EN
where 
\EQ
{d^2\over dt_1^2}D_x(t_1, t_2) = {1\over g_s}\delta(t_1-t_2) . 
\EN
This formal expression 
 is more convenient for comparison with 
Matrix theory. 
It should be noted that the recoil correction is of 
the same order as the 3-body interaction term $S^{(1)}
=\int ds L_3$. 

\subsection{Effective action formalism and recoil in Matrix theory}
We next discuss the treatment of recoil in 
the path integral formalism of Matrix theory. 
Let us first remind ourselves some general  facts 
about the method of effective action. 
To avoid unnecessary 
complexity in the presentation, 
we use simplified notations for the 
products of  correlation functions and their 
functional derivatives, 
suppressing the integrals over the 
time variable. There should not arise any confusion 
for such text-book matters. 

The effective action $\Gamma[\phi]$ is defined by 
\EQ
e^{iW[J]}={\cal N}^{-1}
\int [d\varphi]\, \,  e^{iS[\varphi]-i J\varphi} =
e^{i(\Gamma[\phi]-J\phi)} ,
\EN
\EQ
\phi \equiv \langle \varphi\rangle [J] =
{
\int [d\varphi]\, \,  \varphi \, e^{iS[\varphi]-i J\varphi}
\over 
 \int [d\varphi]\, \,  e^{iS[\varphi]-i J\varphi}}.
\EN
Note that here we discriminate the 
field $\varphi$, which is path-integrated, from its expectation 
value $\phi$. Of course, the field $\varphi$ represents 
all of the different fields of Matrix theory. 
In terms of Feynman diagrams, the effective action 
$\Gamma[\phi]$ is the generating functional  of 
1PI diagrams with their  external lines are given by the 
function $\phi$.  
The familiar properties of the effective action are 
\EQ
{\delta \Gamma[\phi]\over \delta \phi}=J ,
\EN
\EQ
i{\delta \over \delta J}{\delta \Gamma[\phi]\over \delta \phi}=
i{\delta \phi\over \delta J}
{\delta^2 \Gamma[\phi]\over \delta \phi^2}=
\langle \varphi \varphi \rangle_{c} {\delta^2 \Gamma[\phi]\over \delta \phi^2}=i .
\EN
The last equation of course means that the second derivative 
of the effective action is the inverse of the propagator 
(=connected two-point function). 

Assume that the classical action has a solution 
$\phi_0$,
\EQ
{\delta S[\phi]\over \delta \phi}\Big|_{\phi=\phi_0}=0 .
\EN
What we obtain in our eikonal approximation is 
the phase shift defined by 
\EQ
\Delta_{\phi_0}\equiv  \Gamma[\phi_0] , 
\EN
since the calculation amounts to computing 
 the 1PI diagrams by taking the external lines to be 
the straight line trajectories after setting $J=0$.  
Namely, the eikonal effective action is just the 
special value of the exact effective action 
which is evaluated by setting the 
would-be expectation value $\langle \varphi \rangle$ 
to be the classical background. 
Of course, the classical background does  not   
satisfy the variational equation of the exact effective action, 
\EQ
{\delta \Gamma[\phi]\over \delta \phi}\Big|_{\phi=
\phi_0} \ne 0 .
\EN

In the calculation of scattering phase shift, 
the external source term is zero  in the bulk and 
 is only nonzero at infinite 
future and past, leading  to the surface term 
as explained in the beginning of the 
present section. With this understanding, 
we have the identity,  
\EQ
e^{i\Gamma[\phi]}={\cal N}^{-1}
\int [d\tilde{\varphi}]\, \,  e^{iS[\phi_0+\tilde{\varphi}]} ,
\EN
as a consequence of the definition 
of the effective action. 
Here the exact vacuum expectation value, 
\EQ
\phi=\phi_0+ \langle \tilde{\varphi}\rangle ,
\EN
satisfies the quantum equation of motion 
\EQ
{\delta \Gamma[\phi]\over \delta \phi}=0 .
\EN
This shows that obtaining the exact effective action 
for the correct trajectory which satisfies the 
equation of motion, we have to take into account 
the 1P reducible part. The difference between the 
effective action in the eikonal approximation and 
the exact effective action represents the 
recoil effect, 
\EQ
\Gamma_{recoil}[\phi_0] 
\equiv \Gamma[\phi]-\Delta_{\phi_0}  .
\EN
This can be expanded as
\[
{\delta \Gamma[\phi]\over \delta \phi}\Big|_{\phi=\phi_0}
\langle \tilde{\varphi}\rangle 
+{1\over 2}\langle \tilde{\varphi}\rangle{\delta^2 \Gamma[\phi]\over \delta \phi^2}\Big|_{\phi=\phi_0}
\langle \tilde{\varphi}\rangle + \cdots .
\]

The one-point 
function $\langle \tilde{\varphi} \rangle$ is of first order 
with respect to the loop expansion parameter. 
Thus we have the loop expansion as, 
\EQ
\Gamma[\phi]=S[\phi] + \Gamma^{(1)}[\phi] 
+ \Gamma^{(2)}[\phi]+ \cdots ,
\EN
\EQ
\phi=\phi_0 +\phi^{(1)} + \phi^{(2)} \cdots ,
\EN
where the superscript $(i)$ denote the order 
with respect to the loop expansion. 
This leads to  
\EQ
\Gamma_{recoil}[\phi_0]
={\delta \Gamma^{(1)}[\phi]\over \delta \phi}\Big|_{\phi=\phi_0}
 \phi^{(1)}
+{1\over 2}\phi^{(1)}
{\delta^2 S[\phi]\over \delta \phi^2}\Big|_{\phi=\phi_0}
\phi^{(1)} + \cdots .
\EN
Thus the lowest order recoil effect starts from the 
two-loop order. 
On the other hand,  by the definition of the effective action, 
\EQ
{\delta^2 S[\phi]\over \delta \phi^2}=-
{N\over g_s}{d^2\over ds^2} ,
\EN
\EQ
0={\delta \Gamma[\phi]\over \delta\phi}=
{\delta^2 S[\phi]\over \delta \phi^2}\Big|_{\phi=
\phi_0}\phi^{(1)} + {\delta\Gamma^{(1)}[\phi]\over \delta\phi}\Big|_{\phi=\phi_0} + \cdots .
\EN
Then, the final form of the lowest order recoil 
effect is expressed as 
\EQ
\Gamma_{recoil}^{(2)}[\phi_0]=
i{1\over 2}{\delta \Gamma^{(1)}[\phi]\over \delta \phi}\Big|_{\phi=\phi_0}\langle 
\tilde{\varphi}\tilde{\varphi} \rangle_0 
{\delta \Gamma^{(1)}[\phi]\over \delta \phi}\Big|_{\phi=\phi_0}
=
{1\over 2}{g_s\over N}{\delta \Gamma^{(1)}[\phi]\over \delta \phi}\Big|_{\phi=\phi_0}\Bigl({d\over ds}\Bigr)^{-2} {\delta \Gamma^{(1)}[\phi]\over \delta \phi}\Big|_{\phi=\phi_0} .
\EN
This coincides with the supergravity result 
by identifying the recoil acceleration of the D-particle 
as 
\EQ
\delta \alpha = {g_s\over N} {\delta \Gamma^{(1)}[\phi]\over \delta \phi}\Big|_{\phi=\phi_0} = {d^2\over ds^2} \phi^{(1)}.
\EN
Note again that the recoil correction 
$\Gamma_{recoil}^{(2)}[\phi_0]$ is of the same order 
as the 1PI part $\Gamma^{(2)}[\phi_0]$. 

The above argument shows that in order 
to check the agreement of recoil 
between supergravity and Matrix theory up to two-loop order, 
it is sufficient to compute the one-loop tadpole 
diagrams for the matrix fields. 
The agreement requires that 
the tadpoles  vanish for the gauge 
field and the off-diagonal components of the Higgs fields, 
and that the diagonal components of the Higgs tadpole must 
coincide with the recoil shifts (\ref{shiftgeneral}). 
Since the effective action for the 
diagonal Higgs fields is derived using the eikonal 
approximation, it is not completely tautological 
to check that the tadpole of the diagonal Higgs field 
is given by the variational equation of motion 
derived from our effective action. 
Conversely, the explicit check of these properties 
provides a strong support for our method  
starting from the eikonal approximation.  

\subsection{Matrix-theory calculation}
Leu us now confirm that all the above requirements 
are indeed satisfied in Matrix theory. 
The one-point functions of ghost, anti-ghost 
and fermionic fields trivially vanish 
and that of the gauge field also vanishes
by explicit calculation. 
Taking the background configuration of the Higgs field
as $\delta_{ij} (\vec{\tilde{v}}_i \tau + \vec{x}_i)
\equiv \delta_{ij} \vec{r}_{i}(\tau)$, 
the propagators of all the fluctuating 
fields can be expressed in terms of proper-time 
scalar propagator, 
\EQ
\Delta_{ij} (\sigma,\tau_1,\tau_2) \equiv 
\exp \left[ -\sigma  (-\partial_{\tau_1}^2 + r_{ij}(\tau_1)^2 )
\right] \delta(\tau_1 -\tau_2) , 
\EN
where $\vec{r}_{ij} = \vec{r}_{i} - \vec{r}_{j}$ and 
we used the Euclidean metric.   
To avoid confusion, we note that in this subsection 
 the indices ($i, j, \ldots$) 
are matrix indices.  The blocks with equal 
velocities and initial positions on the other 
hand are denoted by using the 
indices ($a, b, \ldots$). 
The only nonvanishing one-point function is that of diagonal components
of the fluctuating Higgs field $Y_{ii}$.  
We found
\begin{eqnarray}
\langle \vec{Y}_{ii} (\tau) \rangle =
g_s \int_{-\infty}^{\infty} d \tau'
\Delta_0 (\tau-\tau')
\sum_{j} \int_0^\infty d \sigma
\left[
-32 \tilde{V}_{ij} (\sigma)^4 \vec{r}_{ij}(\tau')
\Delta_{ij} (\sigma,\tau',\tau') 
\qquad \quad \right. \nonumber \\ \left.
-32 \tilde{C}_{ij} (\sigma) \tilde{V}_{ij} (\sigma)^2 
\vec{\tilde{V}}_{ij} (\sigma) 
\partial_{\tau'} \Delta_{ij} (\sigma,\tau',\tau')
\right] 
+ O(g_s^2),
\label{onepoint}
\end{eqnarray}
where   
\begin{eqnarray}
\vec{\tilde{v}}_{ij} \equiv
\vec{\tilde{v}}_{i} - \vec{\tilde{v}}_{j}, \quad
\vec{\tilde{V}}_{ij} (\sigma) \equiv
\frac{\vec{\tilde{v}}_{ij}}{\tilde{v}_{ij}} 
\sinh \frac{\sigma \tilde{v}_{ij}}{2}, \quad
\tilde{C}_{ij} (\sigma) \equiv
\cosh \frac{\sigma \tilde{v}_{ij}}{2},
\end{eqnarray}
and $\Delta_0 (\tau-\tau')$ satisfies
\begin{eqnarray}
- \partial_{\tau_1}^2 \Delta_{0} (\tau_1-\tau_2) 
= \delta (\tau_1 -\tau_2). 
\end{eqnarray}

The recoil acceleration 
\begin{eqnarray}
\langle \partial_{\tau}^2 \vec{Y}_{ii} (\tau) \rangle =
g_s \sum_{j} \int_0^\infty d \sigma
\left[
32 \tilde{V}_{ij} (\sigma)^4 \vec{r}_{ij} (\tau)
\Delta_{ij} (\sigma,\tau,\tau) 
\qquad \quad \right. \nonumber \\ \left.
+32 \tilde{C}_{ij} (\sigma) \tilde{V}_{ij} (\sigma)^2 
\vec{\tilde{V}}_{ij} (\sigma) 
\partial_{\tau} \Delta_{ij} (\sigma,\tau,\tau)
\right] 
+ O(g_s^2),
\end{eqnarray}
is local with respect to time and its leading contribution
with respect to the relative velocities is
\begin{eqnarray}
\langle \partial_{\tau}^2 \vec{Y}_{ii} (\tau) \rangle_{\rm leading} =
g_s \sum_{j} \left[
\frac{105}{16} \frac{\tilde{v}_{ij}^4 \vec{r}_{ij} (\tau) }
{r_{ij} (\tau)^9}
+\frac{15}{4} \tilde{v}_{ij}^2 \vec{\tilde{v}}_{ij}
{d\over d\tau} \frac{1}{r_{ij} (\tau)^7}
\right]
+ O(g_s^2),
\end{eqnarray}
which is in Minkowski metric ($\tau\rightarrow is$),
\begin{eqnarray}
\langle \partial_s^2 \vec{Y}_{ii} (s) \rangle_{\rm leading} =
- g_s \sum_{j} \left[
\frac{105}{16} \frac{v_{ij}^4 \vec{r}_{ij} (s) }
{r_{ij} (s)^9}
+\frac{15}{4} v_{ij}^2 \vec{v}_{ij}
{d\over ds}\frac{1}{r_{ij} (s)^7}
\right]
+ O(g_s^2).
\end{eqnarray}
After rewriting the sum over the diagonal 
indices ($i, j, \ldots$) to that ($a, b, \ldots$) over the 
diagonal blocks corresponding to the 
clusters of D-particles, 
this is precisely the form we have derived
 from the effective Lagrangian.

\vspace{0.3cm}
\section{Concluding Remarks} 
We have completed the comparison of 
11 dimensional supergravity in its classical 
approximation and Matrix theory up to 
two-loop order by properly taking into 
account the recoiled motion of D-particles. 
There are many directions to extend our  works in I and 
the present paper.   
We enumerate some of them. 
\begin{enumerate} 
\item Investigate the case with  nontrivial 
gravitational background fields. 
On Matrix-theory side, it is not at all clear how to 
treat nontrivial background fields. 
At least some parts of the gravitational 
degrees of freedom are contained in the 
off-diagonal part of the matrix fields. 
Deeper understanding of the dynamics of 
off-diagonal part might be useful to answer 
the problem pointed out in \cite{douglas}. 
\item Extend the computation beyond 
classical approximation. It is not obvious 
how to take into account the higher-dimension 
effects \cite{green} on supergravity side. 
 From the viewpoint of 
type IIA superstring, this amounts to 
considering the higher effects in $\alpha'$ and genus-expansions. 
On Matrix theory side, such higher order effects are 
contained in the subleading contributions  
with respect to the velocity $\alpha' v/r^2$ 
and also to the $1/N$ expansion.  
To clarify the situation of higher order effects is 
important to deal with the question posed in \cite{vkbb}. 
We have to discriminate two different origins 
of higher dimension effects related to the 
string extension and string loop effects. 
It should also be kept in mind that in going 
to higher orders corresponding to quantum loops of 
supergravity, the subtlety of the 
zero modes \cite{ph} might become really relevant. 
\item Extend the comparison between 
supergravity and Matrix theory to other higher-dimensional branes.  For such examples, see ref. \cite{dan}. 
It is also important to include the spin effect \cite{spin} 
and make 
supersymmetric completion \cite{susy} of the effective multi-body 
interactions. 
\end{enumerate}

\vspace{0.4cm}
\noindent
Acknowledgements

We would like to thank M. Ikehara and Y. Kazama for valuable discussions.  The work of Y. O. is supported 
in part by the Japan Society for the Promotion of 
Science under the Predoctoral Research Program (No. 08-4158). The work of T.Y. is supported in part 
by Grant-in-Aid for Scientific  Research (No. 09640337) 
and Grant-in-Aid for International Scientific Research 
(Joint Research, No. 10044061) from the Ministry of  Education, Science and Culture.

\end{document}